\documentclass[prl,showpacs,superscriptaddress,twocolumn]{revtex4-1}
\usepackage{latexsym}
\usepackage{amsfonts}
\usepackage{amssymb}
\usepackage{amsmath}
\usepackage{graphicx}

\begin{document}

\title{Magnetic domain wall motion by spin transfer - D\'eplacement de paroi magn\'etique par transfert de spin}
 
\author{J. Grollier}
\affiliation{Unit\'e Mixte de Physique CNRS/Thales and Universit\'e Paris Sud 11, 1 ave A. Fresnel, 91767 Palaiseau, France}
\author{A. Chanthbouala}
\affiliation{Unit\'e Mixte de Physique CNRS/Thales and Universit\'e Paris Sud 11, 1 ave A. Fresnel, 91767 Palaiseau, France}
\author{R. Matsumoto}
\affiliation{Unit\'e Mixte de Physique CNRS/Thales and Universit\'e Paris Sud 11, 1 ave A. Fresnel, 91767 Palaiseau, France}
\author{A. Anane}
\affiliation{Unit\'e Mixte de Physique CNRS/Thales and Universit\'e Paris Sud 11, 1 ave A. Fresnel, 91767 Palaiseau, France}
\author{V. Cros}
\affiliation{Unit\'e Mixte de Physique CNRS/Thales and Universit\'e Paris Sud 11, 1 ave A. Fresnel, 91767 Palaiseau, France}
\author{F. Nguyen van Dau}
\affiliation{Unit\'e Mixte de Physique CNRS/Thales and Universit\'e Paris Sud 11, 1 ave A. Fresnel, 91767 Palaiseau, France}
\author{A. Fert}
\affiliation{Unit\'e Mixte de Physique CNRS/Thales and Universit\'e Paris Sud 11, 1 ave A. Fresnel, 91767 Palaiseau, France}

\begin{abstract}
The discovery that a spin polarized current can exert a large torque on a ferromagnet through a transfusion of spin angular
momentum, offers a new way to control a magnetization by simple current injection, without the help of an applied external field. Spin transfer can be used to induce magnetization reversals and oscillations, or to control the position of a magnetic domain wall. In this review, we focus on this last mechanism, which is today the subject of an extensive research, both because the microscopic details for its origin are still debated, but also because promising applications are at stake for non-volatile magnetic memories.

\vspace{10 pt}

Par transfert de moment cin\'etique, un courant polaris\'e en spin peut exercer un couple sur l'aimantation d'un nano-aimant. Cette d\'ecouverte permet de contr\^oler la direction d'une aimantation par simple injection de courant, sans l'aide d'un champ magn\'etique ext\'erieur. Le transfert de spin peut \^etre utilis\'e pour induire des renversements ou des oscillations d'aimantation, ou encore pour contr\^oler la position d'une paroi magn\'etique. Dans cette revue, nous nous concentrerons sur ce dernier m\'ecanisme, qui est aujourd'hui le sujet d'intenses recherches. En effet, non seulement ses origines microscopiques sont encore sujettes \`a d\'ebat, mais de plus de tr\`es prometteuses applications aux m\'emoires magn\'etiques non-volatiles sont en jeu. 
\end{abstract}

\maketitle

\large\textbf{1-Introduction on the spin transfer effect}

\vspace{10 pt}

The spin transfer effect allows to manipulate the magnetization of a nanomagnet without the help of an applied magnetic field. This phenomena was theoretically predicted in 1996 by L. Berger \cite{Berger:PRB:1996} and J. Slonczewski \cite{Slonczewski:JMMM:1996}. It immediately attracted a lot of attention both for its fundamental interest as a new spintronic effect and its huge potential for applications. 
The spin transfer takes its origin in the transfusion of magnetic momentum from a spin polarized current to the local magnetization. For large current densities, typically of the order of 10$^7$ A.cm$^{-2}$, the spins carried by the conduction electrons can exert a torque large enough to reverse or destabilize the magnetization of a small magnetic object, typically with lateral dimensions smaller than a few hundred nanometers. 

\begin{figure}
	\includegraphics[width=.45\textwidth]{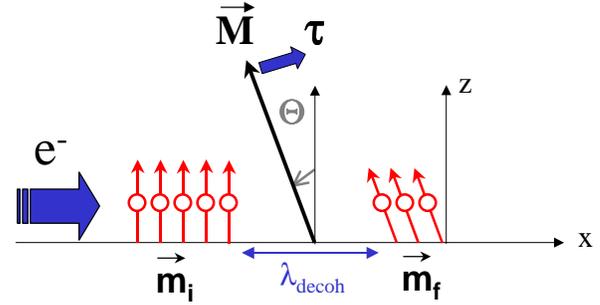}
	\caption{Illustration of the spin transfer effect principle. Here, the spin transfer torque $\mathbf{\tau}$ induced by the spin-polarized current tends to tilt the magnetization $\mathbf{M}$ towards the z axis.}
	\label{fig1}
\end{figure}	

The principle of spin transfer is illustrated on Fig.\ref{fig1}. A spin polarized current, with spins pointing up in the z direction is propagating in the x direction. It enters a ferromagnetic zone, with the magnetization $\mathbf{M}$ canted to an angle $\theta$ with respect to z. The spin current therefore has a component transverse to the local magnetization. Due to the large exchange field ($\approx$ 1000 T), the spins carried by the conduction electrons start to precess around $\mathbf{M}$. After approximately one rotation the coherence is lost due to the diffuse transport in ferromagnetic metals, and in average, the spins are aligned with $\mathbf{M}$ : the transverse spin component has been lost. Through the conservation of momentum, it is transfered to the magnetization $\mathbf{M}$ in the form of a torque. This torque tends to rotate the magnetization vector, in the case illustrated on Fig.\ref{fig1} towards the z axis. The direction of this spin transfer torque is related to the sign of the injected current. Its amplitude depends on the degree of polarization of the spin current, as well as the current density. The decoherence process leading to the loss of the transverse component of the spin current is practically an interface effect \cite{StilesMilat}. It occurs on a very small lengthscale $\lambda_{decoh}$, typically for 3d transition metals like Co, $\lambda_{decoh}$ $\approx$ 1-2 nm.   

The first experiments clearly proving the existence of the spin transfer effect date from the years 2000 \cite{Tsoi:PRL:1998,Myers:Science:1999,Katine:PRL:2000,Grollier:APL:2001}. Magnetization manipulation by spin transfer can take several forms. It is possible either to switch or drive into sustained oscillations a quasi-uniform magnetization, i.e. a macrospin as already reviewed in this journal \cite{Cros:CRAS:2005}, or it is possible to control the position of a magnetic domain wall.

We will concentrate in this paper on spin transfer induced magnetic domain wall motion.
Magnetic domain walls have always attracted a lot of studies, for fundamental and applicative reasons. They are small magnetic objects (a Bloch wall width can be below 10 nm), that propagate with large speeds \cite{Hayashi:PRL:2007,Pizzini:APEX:2009,Ulhir:PRB:2010} ($\approx$ 100 m.s$^{-1}$), and can therefore be used to transmit or store information. 

In section 2, we will present the first experiments that proved the possibility to manipulate a magnetic domain by spin transfer. In section 3, we will briefly describe the theoretical background for current-induced domain wall motion in the classical geometry where the current is injected in the plane of the magnetic layers. Section 4 will review the most important applications foreseen for this phenomena, with a special focus on the most recent : the spintronic memristor. This will lead us to the last section (5), dedicated to recent developments showing that another geometry, using perpendicular current injection, opens new exciting perspectives to spin transfer induced domain wall motion. 

\vspace{10 pt}

\large\textbf{2-The first experiments}

\vspace{10 pt}

As soon as in the 80s, Luc Berger has the intuition that a spin polarized current can interact with a magnetic domain wall, leading to its motion. He very early proposed different physical mechanisms that could potentially lead to such effect \cite{Berger:JAP:1978,Berger:JAP:1984,Berger:JAP:1988}. These studies converged toward the spin transfer theory in 1996 \cite{Berger:PRB:1996,Slonczewski:JMMM:1996}.  Together with his team, he also performed pioneer experiments to put this interaction into evidence \cite{Freitas:JAP:1985,Hung:JAP:1988,Hung:JAP:1990}. Nevertheless in the 80s, the lithography techniques allowing to fabricate samples with sub-micrometer dimensions were not available. Two main problems were therefore faced. First, it was not possible to isolate a single domain wall. Secondly, the macroscopic dimensions of the samples prevented the injection of large current densities without overheating. That's why the first conclusive experiments were only performed in the years 2000 \cite{Grollier:JAP:2002,Klaui:APL:2003,Grollier:APL:2003}. By using e-beam lithography techniques, it was then possible to fabricate magnetic stripes, a few micrometers long, with a cross section of a few nm by a few hundred nm. This stripe geometry favours magnetization reversal by domain wall nucleation and propagation. By engineering domain wall traps, using for example constrictions in the stripe, or simply the natural defects arising from the lithography process, it was possible to pin a single magnetic domain wall. Thanks to the small cross section of the samples, the current densities necessary to move the domain wall where reached for reasonable injected dc currents of typically a few mA. The domain wall position was detected by transport measurement. For stripes made of a single magnetic material, the Anisotropic Magnetoresistance (AMR) was used \cite{Klaui:APL:2003}.  Nevertheless, AMR has two disavantages, first it will always be a small effect ($<$ 1 $\%$), secondly it allows only to detect the presence of a DW, not its position. 
 
\begin{figure}
	\includegraphics[width=.45\textwidth]{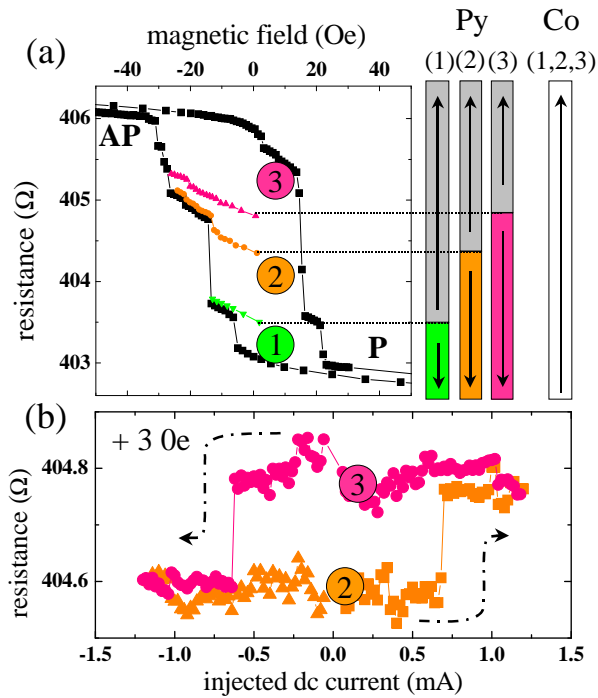}
	\caption{(a) GMR minor cycle associated with the reversal of the permalloy layer of the Co/Cu/NiFe trilayer at T$=$300 K. The field is applied along the stripe. The magnetization of the Co layer is pinned in the positive field direction. Colored curves : variation of the resistance when the cycle is stopped at one of the plateaus and the field is brought back to zero. Also sketched are the DW position in the Py stripe and the magnetic configuration corresponding to the levels 1, 2, and 3. (b) Resistance vs current in very low constant field H (3 Oe) along the stripe : motion from 2 to 3 with
a positive current and back to 2 with a negative current. The numbers 2 and 3 refer to the DW configurations and corresponding resistance levels of
(a). A small contribution ($\propto$ I$^2$), due to the joule heating ($\Delta$T $\approx$ 5 K), has been subtracted for clarity.}
	\label{fig2}
\end{figure} 

In our experiments, we used a different detection scheme \cite{Grollier:JAP:2002,Grollier:APL:2003}. Our stripes were composed of a magnetic metal 1 / normal metal / magnetic metal 2 sandwich, called spin-valve. This allowed us to use the Giant Magnetoresistance effect (GMR) \cite{Baibich:PRL:1988} (that can reach more than 10 $\%$ at room temperature) to detect the domain wall position. We chose a highly coercive Cobalt layer for the bottom magnetic layer 2, so that its magnetization would remain fixed. For the top layer on the contrary we used a NiFe layer with low coercivity, in which a domain wall could easily propagate. Due to the GMR effect, when the magnetizations of the Co and NiFe layers are parallel (P state), the resistance is low, R$_P$ $\approx$ 402.7 $\Omega$ as can be seen from the resistance versus field curve of Fig. \ref{fig2}(a). When they are antiparallel (AP state), the resistance is larger R$_{AP}$ $\approx$ 406 $\Omega$.  When a domain wall propagates in the NiFe layer, the resistance takes intermediate states between R$_P$ and R$_{AP}$, since $R = R_{AP}.x/L+R_P.(1-x/L)$ where x is the domain wall position and L the length of the wire between the contacts. Due to the imperfect lithography process, the stripe has some edge roughness that naturally tends to pin the domain wall at some specific positions. We first apply a magnetic field along the stripe to control the initial domain wall position. In the resistance versus field curve, the domain wall pinning appears as resistance plateaus, each plateau corresponding to a given pinning site (labelled 1-3 in Fig. \ref{fig2}(a)). In order to study the curent-induced domain wall motion, we trapped the domain at the pinning center 2, by monitoring the resistance with very low applied current. Then we ramped up the dc current. As can be seen from Fig. \ref{fig2}(b), at about + 0.7 mA, a resistance jump occurs, corresponding to domain wall motion to the plateau number 3. By sweeping the current to negative values, the domain wall moved back to its initial position 2 at I $\approx$ - 0.7 mA. This hysteresis cycle proves the possibility to induce back and forth domain wall motion by current injection. At the low fields (3 Oe) at which this experiment was performed, the domain wall propagation direction depends on the sign of the current, which is consistent with spin transfer effects (see Fig. \ref{fig1}). Because the stripe is not composed of a single material, the exact determination of the current density through the NiFe layer in which the domain wall propagates is delicate. We have estimated that a higher limit for the current density corresponding to the threshold current 0.7 mA is 9.3 10$^6$ A.cm$^{-2}$, in agreement with the predicted current densities \cite{Berger:PRB:1996,Slonczewski:JMMM:1996}.

\vspace{10 pt}

\large\textbf{3-Theory for domain wall motion in the classical lateral geometry}

\vspace{10 pt}

The first experiments were performed using magnetic materials in which the magnetization naturally lies in the plane of the layers (in-plane magnetic anisotropy). NiFe is still the most widely used material thanks to its low coercivity. In this case, the stripe geometry favours the formation of Neel type walls. The magnetization rotates in the plane of the layer, resulting in domain wall sizes of a few hundred nanometer, comparable to the stripe width. When a current is injected in the stripe, it gets polarized by s-d interaction with the local magnetization. Since the wall width is much larger than the spin-diffusion length ($\approx$ 4 nm at room temperature in NiFe) or the coherence length $\lambda_{decoh}$ ($\approx$ 1-2 nm), the magnetization rotates slowly enough for the spins of the conducting electrons to follow adiabatically the local spins in the wall. The spin-transfer torque can then be written for the continuously rotating magnetization $\mathbf{m}$ in the wall : $\mathbf{T_{STT}} = - (\textbf{u}.\mathbf{\nabla})\mathbf{m}$, where  $\mathbf{u}$ is a velocity proportional to the amplitude of the torque : 
$\mathbf{u}=JPg\mu_B/(2eM_s)$, where J is the current density, P the spin polarization, g the g-factor, $\mu_B$ the Bohr magneton, e the electron charge, and $M_s$ the magnetization at saturation of the ferromagnetic layer. $\mathbf{m}$ is a unit vector along the magnetization.

\begin{figure}
	\includegraphics[width=.45\textwidth]{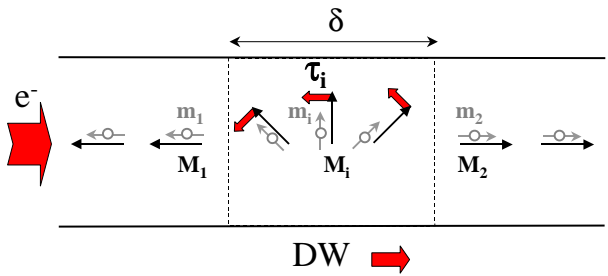}
	\caption{Sketch of the spin transfer ballistic torque on a domain wall. }
	\label{fig3}
\end{figure}	

In order to examine the consequences of the spin transfer torque on the magnetization motion, it is useful to add it into the Landau-Lifshitz-Gilbert equation that describes the magnetization dynamics : 

\begin{equation}\label{LLG1}
\frac{d\mathbf{m}}{dt} = - \gamma_0 \mathbf{m} \times \mathbf{H_{eff}} + \alpha \mathbf{m} \times \frac{d\mathbf{m}}{dt} - (\mathbf{u}.\mathbf{\nabla})\mathbf{m}
\end{equation} 

In this equation, $\gamma_0$ is the gyromagnetic ratio, $\mathbf{H_{eff}}$ the effective field including the applied magnetic field but also other contributions from the different magnetic anisotropies and $\alpha$ is the viscous Gilbert damping.
Both by solving analytically this equation or by introducing it in micromagnetic simulations, the predicted threshold current densities for domain wall motion were found one order of magnitude too large compared to experiments \cite{Thiaville:JAP:2004}. Thiaville \textit{et al.} soon pointed out that this discrepancy arises from the fact that the spin transfer torque as introduced in Eq. (\ref{LLG1}) does not have the proper symmetry to push the domain wall \cite{Thiaville:EP:2005} . Indeed efficient domain wall motions are obtained by a rotation of the spins around the large demagnetizing field. This demagnetizing field originates from the magnetic charges created when the spins in the DW are tilted out-of-plane. The spin transfer torque written in the form of Eq. (\ref{LLG1}) is in plane, and therefore cannot induce this demagnetizing field. Different authors then suggested that in order to understand the experimental current densities, it was necessary to introduce an additional term in the LLG equation \cite{Thiaville:EP:2005,Zhang:PRL:2004}:

\begin{eqnarray}\label{LLG2}
\frac{d\mathbf{m}}{dt} = - \gamma_0 \mathbf{m} \times \mathbf{H_{eff}} &+& \alpha \mathbf{m} \times \frac{d\mathbf{m}}{dt} - (\mathbf{u}.\mathbf{\nabla})\mathbf{m} \nonumber \\
&+& \beta \mathbf{m} \times \left[(\mathbf{u}.\mathbf{\nabla})\mathbf{m}\right]
\end{eqnarray} 

This last torque points out-of-plane, and generates in turn the demagnetizing field necessary for efficient wall motion. It has the same symmetry as a magnetic field applied in the direction of one of the magnetic domains sandwiching the wall. $\beta$ is like the Gilbert damping $\alpha$ a dimensionless parameter. If the necessity to introduce this beta term is now widely recognized, its physical origin, as well as its amplitude compared to $\alpha$ is still under debate. 

For very narrow domain walls as can be obtained for perpendicular anisotropy materials such as FePt, CoPt or CoNi, a term with the symmetry of $T_{\beta}$ naturally arises from the interaction of the current with the wall. Indeed, when the wall thickness is limited to a few nm, the condition of adiabacity is not fulfilled anymore. Tatara and Kohno have calculated that in this case, the spin transfer effect becomes weak and the dominant contribution is momentum transfer, due to the reflection of electrons on the wall \cite{Tatara:PRL:2004}. 

Nevertheless, this non-adiabatic torque is basically zero for the case of thick magnetic walls, and cannot systematically be identified to the last term of Eq. (\ref{LLG2}) \cite{Xiao:PRB:2006}. The most recent theories show that, in fact, in the adiabatic limit, spin relaxation processes that are at the origin of the damping $\alpha$ (spin-flip scattering or spin-orbit coupling) also induce a second term, similar to $\beta$, when the transport is taken into account \cite{Kohno:JPSJ:2006,Tserkovnyak:PRB:2006,Garate:PRB:2009}. Nevertheless, there is no consensus on the value of $\beta$, some calculations indicating that $\beta$ should be equal to $\alpha$ \cite{Tserkovnyak:PRB:2006}, while other models point our that this is not the case in general \cite{Kohno:JPSJ:2006,Garate:PRB:2009}.

The experimental studies also give very disperse values for the $\beta$ parameter. For vortex walls in Permalloy, $\beta$ has been determined by different methods : imaging and measuring the current-induced wall displacements ($\beta$ $\approx$ $\alpha$) \cite{Meier:PRL:2007}, current pulse induced oscillatory DW depinning ($\beta$ $\approx$ 8 $\alpha$ in ref \cite{Thomas:Nature:2006} and $\beta$ $\approx$ 2 $\alpha$ in ref \cite{Hayashi:APL:2008}), field-assisted current-induced domain wall depinning ($\beta$ $\approx$ 2 $\alpha$) \cite{Lepadatu:PRL:2009}. This diversity of results also occurs for perpendicular anisotropy materials with sharp domain walls as Co/Pt, where a $\beta$ value of 0.35 has been measured by field-assisted current-induced domain wall depinning (to be compared to $\alpha$ $\approx$ 0.15) \cite{Boulle:PRL:2008}, and $\beta$ $\leq$ 0.02 by comparing current and field slight domain walls displacement in a potential well \cite{Miron:PRL:2009}.

This dispersion of results most probably originates from additional contributions to the spin transfer torque, varying depending on the method employed to determine $\beta$. Among them can be listed the current-induced Oersted field \cite{Ulhir:PRB:2010}, Joule heating, all spin-orbit phenomena including Rashba effect \cite{Miron:Nature:2010}, automotive force \cite{Chauleau:PRB:2010}.

Recent experiments use thermally activated domain wall depinning with sub-threshold current densities in order to determine the $\beta$ parameter \cite{Burrowes:Nature:2009,Eltschka:PRL:2010}. This method has the advantage to work in the linear regime, and also to directly integrate thermal effects otherwise regarded as undesirable. In this way, Burrowes \textit{et al.} \cite{Burrowes:Nature:2009} find values of $\beta$ slightly smaller than $\alpha$ ( $\beta_{CoNi}$ = 0.022 to be compared to $\alpha_{CoNi}$ = 0.032; and $\beta_{FePt}$ = 0.06 to be compared to $\alpha_{FePt}$ = 0.1) in perpendicularly magnetized materials. These low values are very surprising since the sharp domain walls should give rise to a non-negligible non-adiabatic contribution to $\beta$, in addition to the spin relaxation adiabatic term. By a similar method Eltschka \textit{et al.} \cite{Eltschka:PRL:2010} determine in Permalloy that $\beta$ = 0.01 for a Transverse Wall, and 0.073 for a Vortex Wall. The larger value for the vortex wall is attributed to a larger non-adiabatic contribution due to the large gradient of magnetization in the vicinity of the vortex core. This last result evidences that the $\beta$ term is not intrinsic to the material, but is highly sensitive to the micromagnetic structure of the wall, which is another point explaining the dispersion of results for a given material.  

\vspace{10 pt}

\large\textbf{4-Applications}

\vspace{10 pt}

Several cutting edges devices using the spin transfer induced DW motion have been proposed. Most of them concentrate on memory effects, taking advantage of the non-volatility of magnetic domains (storing bits) and the speed of DW propagation (fast writing). In the following, we will review the two main applications called the "`racetrack memory"' and the DW-RAM (Random Access Memory). Then we will describe a new emerging and exciting potential application of current-induced DW motion to neuromorphic systems : the DW spintronic memristor.

\begin{figure}
	\includegraphics[width=.45\textwidth]{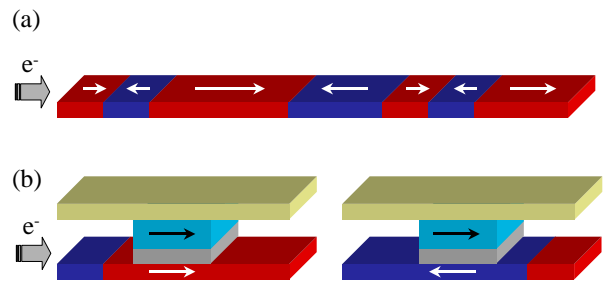}
	\caption{Illustration of the principles of (a) the racetrack memory (b) the DW-RAM}
	\label{fig4}
\end{figure}	

\textsc{\textbf{Racetrack memory}}
 
The "`Racetrack Memory"' concept was introduced by S.S.P. Parkin from the Almaden I.B.M. center \cite{Parkin:Science:2008}. The principle is illustrated on Fig. \ref{fig4} (a). The basic element of the racetrack memory is a sub-micrometer wide magnetic stripe with alternating domains pointing in opposite directions, separated by domain walls. One of the key advantages of current induced DW motion is that the DW propagation direction is independent of the domain wall chirality. For a given current polarity, head to head and tail to tail DWs will move in the same direction. It is therefore possible to push back and forth trains of DWs. This is not the case when an external field is applied. A magnetic field tends to increase the size of the domains pointing in the same direction, which leads to DW annihilation until finally for large enough field the magnetization becomes homogeneous. 
The racetrack memory is a storage device. Like in the low cost hard drives, bits are encoded in the alternative domains directions. Contrarily to the hard drives, there is no rotation of mechanical parts, which promises increased speed.  Pulsed current injection is used to move the domains back and forth in order to intersect with the reading (a magnetoresistive device) and writing (fringing fields with controlled direction) elements. The racetrack is composed of an array of such stripes, stacked either horizontally or vertically. A prototype by I.B.M. should be due in 2013.  Nevertheless, one of the major challenges to face for the racetrack memory is the potential degradation due to repeated injection of the large current densities required to move DW ($\approx$ 10$^8$ A.cm$^{-2}$), close to the breakdown electromigration threshold.  

\textsc{\textbf{DW-RAM}}

Another concept of DW based memory is the DW-RAM. The principle is illustrated on Fig. \ref{fig4} (b). Inspired by the architecture of MRAMs (Magnetic Random Access Memory), the building block is a sub-micrometer size magnetic tunnel junction, i.e. a ferromagnetic fixed layer / insulating barrier / ferromagnetic free layer sandwich. This device takes advantage of the large amplitude of the tunnel magnetoresistance effect (TMR), about 100 $\%$ for crystalline MgO tunnel barriers \cite{Yuasa:NatMat:2004,Parkin:NatMat:2004}. As described in our 2004 patent \cite{Cros:Patent:2004}, the idea is to use spin transfer induced domain wall motion to switch the bit from "`0"' (P state) to "`1"' (AP state). 
The free layer is therefore a low coercivity ferromagnetic stripe in which a domain can be moved back and forth depending on the polarity of the pulsed current injected in the stripe, as shown on Fig. \ref{fig4} (b). Recently, the japanese company NEC has built a small prototype of such a DW-RAM \cite{NEC}. They have used perpendicularly magnetized materials (Co/Ni for the free and Co/Pt for the fixed layer) to reduce the domain wall size and increase the thermal stability. The high frequency operation of their prototype should enable the DW-RAM to compete with SRAMs.

\textsc{\textbf{DW spintronic memristor}}

\begin{figure}
	\includegraphics[width=.45\textwidth]{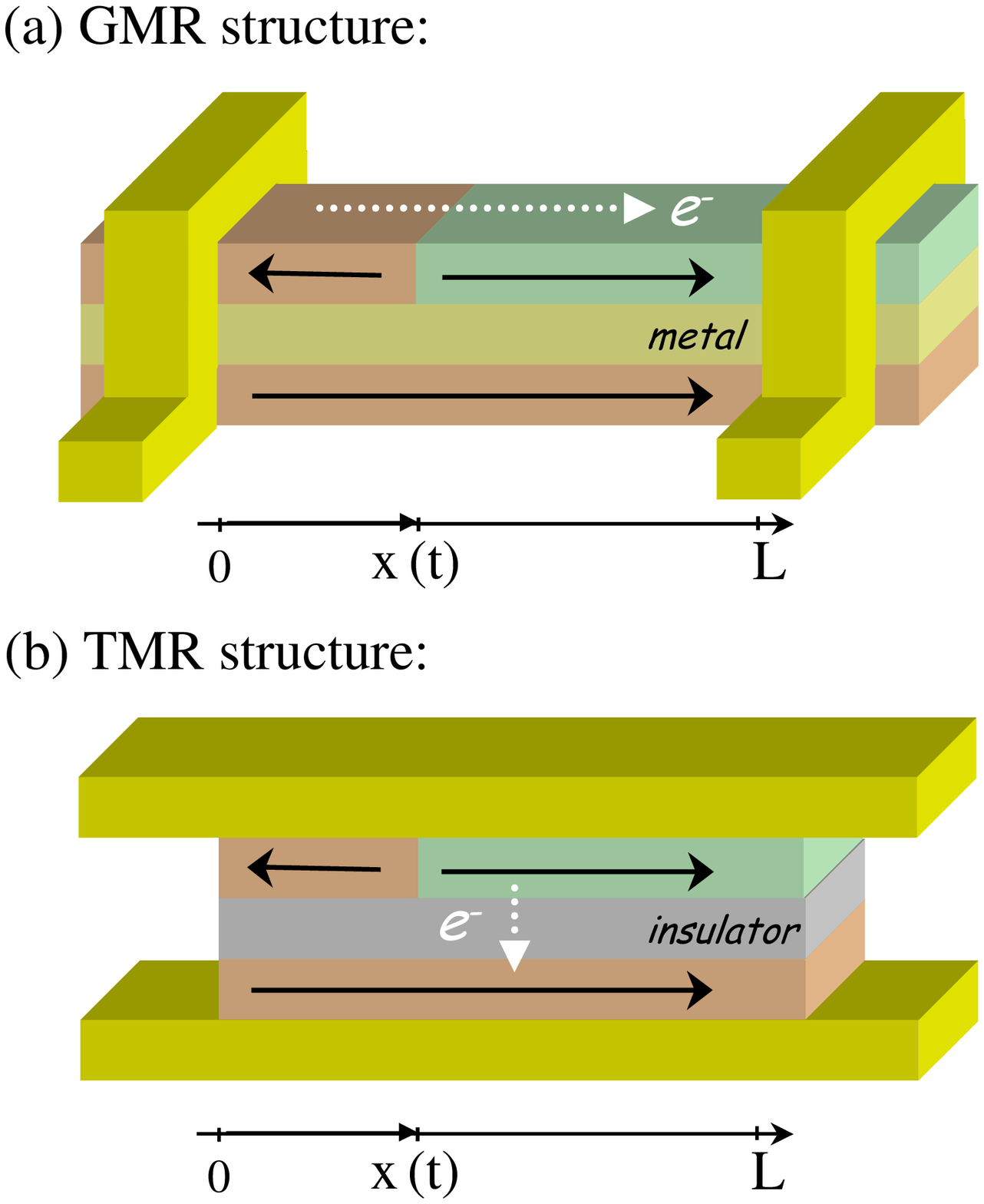}
	\caption{Sketch of the (a) lateral injection GMR  (b) vertical injection TMR spintronic memristors }
	\label{fig5}
\end{figure}

A new exciting potential application of domain wall motion by spin transfer is the spintronic memristor.
The difficulty to provide large numbers of synaptic connections has slowed down the development of neuromorphic circuits, which are still far from the performances of their biological counterparts. Using dynamical analog, reconfigurable nanoscale devices for the synaptic nodes would result in tremendous gains in terms of power, dissipation, miniaturization and computational efficiency . In 2008, the Hewlett-Packard team of S. Williams have demonstrated such devices, called "memristors", that could be the key to future developments of neuromorphic circuits \cite{Strukov:Nature:2008}.

A " memristor " is an analog tuneable resistance. The more intense is the current through the structure, and the longer it is injected, the more the resistance changes . Actually, a memristor is a continuously tuneable resistance M such that $v = M(q) i$ \cite{Chua:IEEE:1971}. This current-voltage relation directly implements the fact that the synapse transmission depends on the information it has previously processed (plasticity). The device proposed by Hewlett-Packard is based on resistance changes due to voltage induced oxygen vacancies displacement in a titanium dioxide layer \cite{Yang:Nature:2008}. Being based on ions electromigration, this device could suffer from potential fragility. 

Spin-transfer induced domain wall motion in a spin valve structure is intrinsically a memristive effect. As we have seen in a previous section, the resistance of the GMR device shown in Fig. \ref{fig5} (a) is $R = R_{AP}.x/L+R_P.(1-x/L)$ where x is the domain wall position and L the length of the wire between the contacts. For current densities above a threshold value Jc, defined in particular by the initial pinning of the DW, the DW propagation speed $u$ is proportional to the injected current : $u = \gamma I$  \cite{Yamaguchi:PRL:2004,Yamanouchi:PRL:2006}. Spin-transfer can generate domain wall speeds above 100 m/s \cite{Hayashi:PRL:2007,Pizzini:APEX:2009}, which means that the resistance of a sub-micron "spin memristor" can be modified in a few nanoseconds. The displacement is given by  : $x(t) = \gamma I t = \gamma q$, where q is the total injected charge. The device resistance depends on the charge, and not only on the current, which confers this memristor its memory effect. For a perfect sample (no pinning center), the memristance is equal to (case of figure 1 (a)) : $M(q)= R_{P}+(R_{AP} - R_P)(\gamma /L) q$. This device is a multi-state analog resistance controlled by the injected charge, via spin transfer induced DW motion. When the current is set back to zero, the device keeps its last resistance value. By changing the initial magnetic configuration, the memristor can be changed from an "inhibitory" (the resistance increases) to an "excitatory" (the resistance decreases) artificial synapse. 

The problem with the device presented above is the small resistance variations obtained when the domain wall is propagating, since the GMR ratio is only a few percent in standard stacks. In order to increase the resistance changes, the tunnel magnetoresistance effect can be used, which requires to replace the metallic normal spacer by a thin insulating barrier, in the manner of the DW-RAM. Nevertheless, the device represented in Fig. \ref{fig4} (b) is not strictly speaking a memristor. Although the resistance can be continuously tuned by current injection through the bottom electrode, it is a three terminal device. The current paths for writing (lateral) and reading (vertical) are not the same and the memristor definition $v = M(q) i$ is not respected. Therefore, fabricating a memristor device using a magnetic tunnel junction would imply to be able to move a domain wall back and forth by vertical current injection, as sketched in Fig. \ref{fig5} (b). This is the mechanism that we propose in a recent patent \cite{Grollier:patent:2010}. In that case, the memristance formula is modified : $M(q)= R_{AP}/[1+(\gamma /L).q.(R_{AP} - R_P)/ R_P]$. The next section is dedicated to the theoretical and experimental grounding of domain wall motion with vertical spin injection. 

\vspace{10 pt}

\large\textbf{5-Domain wall motion by perpendicular spin injection / the domain wall based spintronic memristor}

\vspace{10 pt}

\begin{figure}
	\includegraphics[width=.45\textwidth]{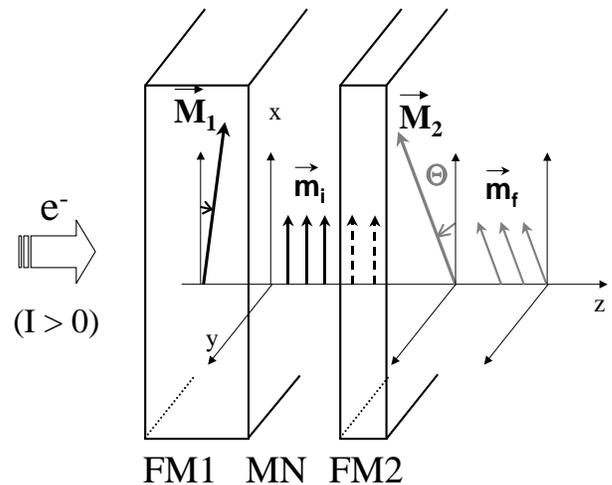}
	\caption{Illustration of the spin transfer effect in a trilayer structure.}
	\label{fig6}
\end{figure}

Vertical current injection through a magnetic trilayer with homogeneous magnetizations is the original case considered by John Slonczewski \cite{Slonczewski:JMMM:1996}. In this type of structure, illustrated in Fig. \ref{fig6}, the spin transfer effect does not arise from a continuously rotating magnetization in a single layer, but from the tilt between two magnetizations in different layers. The reference layer, FM1, is a fixed polarizer for the conduction electrons. The second layer FM2 is the free layer that can be excited by the torque arising from the transfusion of the transverse component of the spin current. In this geometry the spin transfer torque $\mathbf{T_{STT}}$ is composed of two terms $\mathbf{T_{STT}}=\mathbf{T_{IP}}+\mathbf{T_{OOP}}$ \cite{Slonczewski:PRB:2005, Theodonis:PRL:2006}. $\mathbf{T_{IP}}$, the in-plane torque is often referred to as the Slonczewski torque, while $\mathbf{T_{OOP}}$ is indifferently called out-of-plane or field-like torque.
\begin{eqnarray}\label{ST-FLT}
\mathbf{T_{IP}} &=& - \gamma a_J \mathbf{m} \times \left(\mathbf{m} \times \mathbf{m_{ref}} \right)\nonumber \\
\mathbf{T_{OOP}} &=&  - \gamma b_J \mathbf{m} \times \mathbf{m_{ref}}
\end{eqnarray}
$\gamma$ is the gyromagnetic ratio, $\mathbf{m}$ and $\mathbf{m_{ref}}$ are unit vectors along the direction of respectively the free layer and the fixed reference layer, $a_J$ and $b_J$ are the amplitudes of the torques.

Few works have focused on domain wall control by perpendicular current injection \cite{Ravelosona:PRL:2006,Rebei:PRB:2006,Lou:APL:2008,Boone:PRL:2010}. We have recently predicted  \cite{Khvalkovskiy:PRL:2009} that, when the free and reference layers are based on materials with the same anisotropy (either in-plane or perpendicular), the driving torque for domain wall motion is the out-of-plane field-like torque. Indeed, $\mathbf{T_{OOP}}$ produces a magnetic field in the direction of the reference layer, that has the proper symmetry to push the DW along the free layer. On the contrary, the in-plane torque can only shift slightly the domain wall of a few nanometer. While the out-of-plane torque amplitude is very small in metallic spin-valves \cite{Stiles:PRB:2002,Xia:PRB:2002} typically $b_J << 0.1 \: a_J$, it has been shown experimentally that in magnetic tunnel junctions it can reach 30 $\%$ of the in-plane torque \cite{Sankey:Nature:2007,Kubota:Nature:2007}.

In magnetic tunnel junctions with the same composition for the top and bottom electrodes, the out-of-plane field-like torque exhibits a quadratic dependence with bias \cite{Sankey:Nature:2007,Kubota:Nature:2007}.  In order to obtain inverted DW motions for opposite current directions, some degree of asymmetry has to be introduced. It has been recently shown that by using electrodes with different composition, it is possible to induce a linear dependence of the out-of-plane torque at small bias \cite{Oh:Nature:2009,Tang:PRB:2010}.

\begin{figure}
	\includegraphics[width=.45\textwidth]{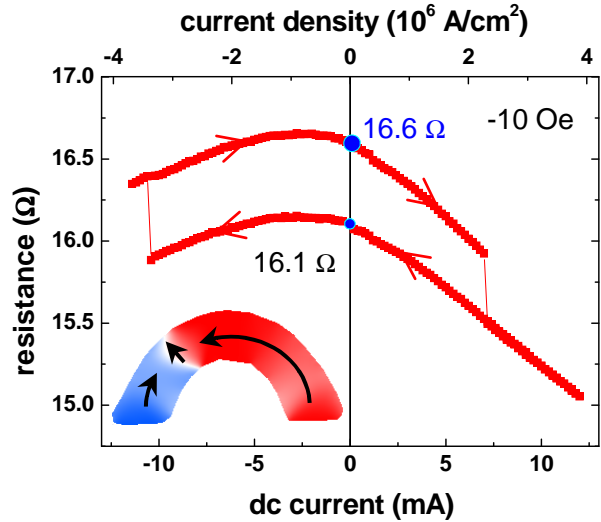}
	\caption{Resistance versus current hysteresis cycle obtained by vertical current-induced domain wall motion in a magnetic tunnel junction. A fixed applied magnetic field of -10 Oe is applied. Inset : micromagnetic simulations (top view of the tunnel junction cross section) showing the initial domain wall position and configuration.}
	\label{fig7}
\end{figure}	

In order to demonstrate the domain wall motion by vertical spin injection, we use a magnetic tunnel junction with a thin MgO barrier (1.1 nm thick), a top (CoFe 1nm/NiFe 4 nm) free electrode and a CoFeB 3 nm reference electrode. The tunnel junction cross section has a specific U-shape, with a wire width of 210 nm. This geometry facilitates DW creation close to the wire edge, as shown by the micromagnetic simulations in the inset of Fig. \ref{fig7}.  In our convention, a positive current corresponds to electrons flowing from the reference to the free layer. In Fig.\ref{fig7} , we show the resistance versus current curve obtained with vertical current injection. The initial resistance is 16.6 $\Omega$. In addition to the expected bias dependence of the tunnel resistance, we clearly observe irreversible resistance jumps. By sweeping the current to positive values, the resistance is switched at I = + 7 mA to a lower resistance state corresponding to another domain wall position, stable at zero current, with a low bias resistance of 16.1 $\Omega$. By applying a negative current of -10.4 mA, it is then possible to move back the domain to its initial position. We thus demonstrate the possibility to move a domain wall back and forth between two pinning centers by perpendicular dc current injection. The current densities corresponding to the DW motion are lower than 4 10$^6$ A.cm$^{-2}$, as can be seen from the top x axis of Fig.\ref{fig7}. The use of perpendicular current injection therefore allows to reduce the current densities by a factor 100 compared to the classical lateral current injection. These recent results pave the way towards the implementation of fast and robust spintronic memristors.

\vspace{10 pt}

\large\textbf{Conclusion}

\vspace{10 pt}

Research on current-induced domain wall motion is a very active field and as shown in section 3, the microscopic origin of the effect is still largely debated. Many promising applications are at stake, among them the racetrack memory, the domain wall RAM, and, as proposed recently, the spintronic memristor. One of the latest developments, domain wall control by vertical current injection, promises a reduction of current densities by more than 2 orders of magnitudes. This result should motivate new advances in the field, accelerate the use of spin transfer induced domain wall motion in industrial devices, and maybe open the path to new applications.

\textbf{Acknowledgments} : 
We thank Daniel Lacour, Amir Hamzic and Manuel Munoz for their collaboration at the early stage of these experiments; Alexei Khvalkovskiy and Konstantin Zvezdin for fruitful discussions and micromagnetic simulations; Akio Fukushima and Shinji Yuasa for their collaboration and magnetic tunnel junction fabrication, Giancarlo Faini for spin-valve stripes fabrication. We acknowledge K. Nishimura, Y. Nagamine, H. Maehara and K. Tsunekawa from Canon Anelva for magnetic tunnel junction growth.
Financial support by the CNRS, JSPS Postdoctoral Fellowships for Research Abroad and the European Research Council (Starting Independent Researcher Grant No. ERC 2010 Stg 259068) is acknowledged.


\begin{thebibliography}{30}
\expandafter\ifx\csname natexlab\endcsname\relax\def\natexlab#1{#1}\fi
\expandafter\ifx\csname bibnamefont\endcsname\relax
  \def\bibnamefont#1{#1}\fi
\expandafter\ifx\csname bibfnamefont\endcsname\relax
  \def\bibfnamefont#1{#1}\fi
\expandafter\ifx\csname citenamefont\endcsname\relax
  \def\citenamefont#1{#1}\fi
\expandafter\ifx\csname url\endcsname\relax
  \def\url#1{\texttt{#1}}\fi
\expandafter\ifx\csname urlprefix\endcsname\relax\def\urlprefix{URL }\fi
\providecommand{\bibinfo}[2]{#2}
\providecommand{\eprint}[2][]{\url{#2}}


\bibitem[{\citenamefont{Berger}(1996)}]{Berger:PRB:1996}
\bibinfo{author}{\bibnamefont{Berger}, \bibfnamefont{L.}},
\bibinfo{title}{Emission of spin waves by a magnetic multilayer traversed by a current}
\bibinfo{journal}{\textit{Phys. Rev. B}} \textbf{\bibinfo{volume}{54}},
\bibinfo{pages}{9353-9358} (\bibinfo{year}{1996}).

\bibitem[{\citenamefont{Slonczewski}(1996)}]{Slonczewski:JMMM:1996}
\bibinfo{author}{\bibnamefont{Slonczewski}, \bibfnamefont{J.C.}},
\bibinfo{title}{Current-driven excitation of magnetic multilayers}  
\bibinfo{journal}{\textit{J. Magn. Magn. Mater.}} \textbf{\bibinfo{volume}{159}},
\bibinfo{pages}{L1-L7} (\bibinfo{year}{1996}).

\bibitem[{\citenamefont{Stiles et~al.}(2006)\citenamefont{Stiles, Miltat}}]{StilesMilat}
\bibinfo{author}{\bibnamefont{Stiles}, \bibfnamefont{M.D.}}
\bibnamefont{\&} 
\bibinfo{author}{\bibnamefont{Miltat},~\bibfnamefont{J.}},
\bibinfo{title}{Spin-transfer torque and dynamics}
in : \bibinfo{editor}{Topics in Applied Physics},
\bibinfo{editor}{B. Hillebrands, A. Thiaville }{Eds.},
\bibinfo{booktitle}{Spin dynamics in confined magnetic structures III},
\textbf{\bibinfo{volume}{101}},
\bibinfo{pages}{225-308},
\bibinfo{publisher}{Springer},
(\bibinfo{year}{2006}). 

\bibitem[{\citenamefont{Tsoi et~al.}(1998)\citenamefont{Tsoi, Jansen, Bass, Chiang, Seck, Tsoi, Wyder}}]{Tsoi:PRL:1998}
\bibinfo{author}{\bibnamefont{Tsoi}, \bibfnamefont{M.}}
\bibinfo{author}{\bibnamefont{\textit{el al.}},~\bibfnamefont{}}
\bibinfo{title}{Excitation of a Magnetic Multilayer by an Electric Current}
\bibinfo{journal}{\textit{Phys. Rev. Lett.}} \textbf{\bibinfo{volume}{80}},
\bibinfo{pages}{4281} (\bibinfo{year}{1998}). 

\bibitem[{\citenamefont{Myers et~al.}(1999)\citenamefont{Myers, Ralph, Katine, Louie, Buhrman}}]{Myers:Science:1999}
\bibinfo{author}{\bibnamefont{Myers}, \bibfnamefont{E.B.}}
\bibinfo{author}{\bibnamefont{\textit{el al.}},~\bibfnamefont{}}
\bibinfo{title}{Current-Induced Switching of Domains in Magnetic Multilayer Devices}
\bibinfo{journal}{\textit{Science}} \textbf{\bibinfo{volume}{285}},
\bibinfo{pages}{867} (\bibinfo{year}{1999}). 

\bibitem[{\citenamefont{Katine et~al.}(2000)\citenamefont{Katine, Albert, Buhrman, Myers, Ralph}}]{Katine:PRL:2000}
\bibinfo{author}{\bibnamefont{Katine},~\bibfnamefont{J.A.}},
\bibinfo{author}{\bibnamefont{\textit{el al.}},~\bibfnamefont{}}
\bibinfo{title}{Current-Driven Magnetization Reversal and Spin-Wave Excitations in Co/Cu/Co Pillars}
\bibinfo{journal}{\textit{Phys. Rev. Lett.}} \textbf{\bibinfo{volume}{84}},
\bibinfo{pages}{3149} (\bibinfo{year}{2000}). 

\bibitem[{\citenamefont{Grollier et~al.}(2001)\citenamefont{Grollier, Cros,
Hamzic, George, Jaffres, Fert, Faini, Ben Youssef, Legall}}]{Grollier:APL:2001}
\bibinfo{author}{\bibnamefont{Grollier}, \bibfnamefont{J.}}
\bibinfo{author}{\bibnamefont{\textit{el al.}},~\bibfnamefont{}}
\bibinfo{title}{Spin-polarized current induced switching in Co/Cu/Co pillars}
\bibinfo{journal}{\textit{Appl. Phys. Lett.}} \textbf{\bibinfo{volume}{78}},
\bibinfo{pages}{3663} (\bibinfo{year}{2001}). 

\bibitem[{\citenamefont{Cros et~al.}(2001)\citenamefont{Cros, Boulle, Grollier, 
Hamzic, Munoz, Pereira, Petroff}}]{Cros:CRAS:2005}
\bibinfo{author}{\bibnamefont{Cros},~\bibfnamefont{V.}},
\bibinfo{author}{\bibnamefont{\textit{el al.}},~\bibfnamefont{}}
\bibinfo{title}{Spin Transfer Torque: a new method to excite or reverse a magnetization}
\bibinfo{journal}{\textit{C. R. Physique}} \textbf{\bibinfo{volume}{6}},
\bibinfo{pages}{956} (\bibinfo{year}{2005}). 

\bibitem[{\citenamefont{Hayashi et~al.}(2007)\citenamefont{Hayashi, Thomas, Rettner, Moriya, Bazaliy, Parkin}}]{Hayashi:PRL:2007}
\bibinfo{author}{\bibnamefont{Hayashi}, \bibfnamefont{M.}}
\bibinfo{author}{\bibnamefont{\textit{el al.}},~\bibfnamefont{}} 
\bibinfo{title}{Current Driven Domain Wall Velocities Exceeding the Spin Angular Momentum Transfer Rate in Permalloy Nanowires}
\bibinfo{journal}{\textit{Phys. Rev. Lett.}} \textbf{\bibinfo{volume}{98}},
\bibinfo{pages}{037204} (\bibinfo{year}{2007}). 

\bibitem[{\citenamefont{Pizzini et~al.}(2009)\citenamefont{Pizzini, Uhlir, Vogel, Rougemaille, Laribi, Cros, Jimenez, Camarero, Tieg, Bonet, Bonfim, Mattana, Deranlot, Petroff, Ulysse, Faini, Fert}}]{Pizzini:APEX:2009}
\bibinfo{author}{\bibnamefont{Pizzini}, \bibfnamefont{S.}}
\bibinfo{author}{\bibnamefont{\textit{el al.}},~\bibfnamefont{}}
\bibinfo{title}{High Domain Wall Velocity at Zero Magnetic Field Induced by Low Current Densities in Spin Valve Nanostripes}
\bibinfo{journal}{\textit{Appl. Phys. Exp.}} \textbf{\bibinfo{volume}{2}},
\bibinfo{pages}{023003} (\bibinfo{year}{2009}). 

\bibitem[{\citenamefont{Ulhir et~al.}(2010)\citenamefont{Uhlir, Pizzini, Rougemaille, Novotny, Cros, Jimenez, Faini, Heyne, Sirotti, Tieg, Bendounan, Maccherozzi, Belkhou, Grollier, Anane, Vogel}}]{Ulhir:PRB:2010}
\bibinfo{author}{\bibnamefont{Uhlir},~\bibfnamefont{V.}}
\bibinfo{author}{\bibnamefont{\textit{el al.}},~\bibfnamefont{}}
\bibinfo{title}{Current-induced motion and pinning of domain walls in spin-valve nanowires studied by XMCD-PEEM}
\bibinfo{journal}{\textit{Phys. Rev. B}} \textbf{\bibinfo{volume}{81}},
\bibinfo{pages}{224418} (\bibinfo{year}{2010}). 

\bibitem[{\citenamefont{Berger}(1978)\citenamefont{Berger}}]{Berger:JAP:1978}
\bibinfo{author}{\bibnamefont{Berger}, \bibfnamefont{L.}},
\bibinfo{title}{Low field magnetoresistance and domain drag in ferromagnets}
\bibinfo{journal}{\textit{J. Appl. Phys.}} \textbf{\bibinfo{volume}{49}},
\bibinfo{pages}{2156} (\bibinfo{year}{1978}). 

\bibitem[{\citenamefont{Berger}(1984)\citenamefont{Berger}}]{Berger:JAP:1984}
\bibinfo{author}{\bibnamefont{Berger}, \bibfnamefont{L.}},
\bibinfo{title}{Exchange interaction between ferromagnetric domain wall and electric current in very thin metallic films}
\bibinfo{journal}{\textit{J. Appl. Phys.}} \textbf{\bibinfo{volume}{55}},
\bibinfo{pages}{1954} (\bibinfo{year}{1984}). 

\bibitem[{\citenamefont{Berger}(1988)\citenamefont{Berger}}]{Berger:JAP:1988}
\bibinfo{author}{\bibnamefont{Berger}, \bibfnamefont{L.}},
\bibinfo{title}{Exchange interaction between electric current and magnetic domain wall containing Bloch lines}
\bibinfo{journal}{\textit{J. Appl. Phys.}} \textbf{\bibinfo{volume}{63}},
\bibinfo{pages}{1663} (\bibinfo{year}{1988}). 

\bibitem[{\citenamefont{Freitas et~al.}(1985)\citenamefont{Freitas,Berger}}]{Freitas:JAP:1985}
\bibinfo{author}{\bibnamefont{Freitas}, \bibfnamefont{P.P.}},
\bibnamefont{\&} 
\bibinfo{author}{\bibnamefont{Berger}, \bibfnamefont{L.}},
\bibinfo{title}{Observation of s-d exchange force between domain walls and electric current in very thin Permalloy films}
\bibinfo{journal}{\textit{J. Appl. Phys.}} \textbf{\bibinfo{volume}{57}},
\bibinfo{pages}{1266} (\bibinfo{year}{1985}). 

\bibitem[{\citenamefont{Hung et~al.}(1988)\citenamefont{Hung,Berger}}]{Hung:JAP:1988}
\bibinfo{author}{\bibnamefont{Hung}, \bibfnamefont{C.-Y.}},
\bibnamefont{\&} 
\bibinfo{author}{\bibnamefont{Berger}, \bibfnamefont{L.}},
\bibinfo{title}{Exchange forces between domain wall and electric current in permalloy films of variable thickness}
\bibinfo{journal}{\textit{J. Appl. Phys.}} \textbf{\bibinfo{volume}{63}},
\bibinfo{pages}{4276} (\bibinfo{year}{1988}).

\bibitem[{\citenamefont{Hung et~al.}(1990)\citenamefont{Hung,Berger,Shih}}]{Hung:JAP:1990}
\bibinfo{author}{\bibnamefont{Hung}, \bibfnamefont{C.-Y.}},
\bibinfo{author}{\bibnamefont{Berger}, \bibfnamefont{L.}},
\bibnamefont{\&} 
\bibinfo{author}{\bibnamefont{Shih}, \bibfnamefont{C.Y.}},
\bibinfo{title}{Observation of a current-induced force on Bloch lines in Ni-Fe thin films}
\bibinfo{journal}{\textit{J. Appl. Phys.}} \textbf{\bibinfo{volume}{67}},
\bibinfo{pages}{5941} (\bibinfo{year}{1990}).

\bibitem[{\citenamefont{Grollier et~al.}(2002)\citenamefont{Grollier, Lacour, Cros,
Hamzic, Vaures, Fert, Adam, Faini}}]{Grollier:JAP:2002}
\bibinfo{author}{\bibnamefont{Grollier}, \bibfnamefont{J.}}
\bibinfo{author}{\bibnamefont{\textit{el al.}},~\bibfnamefont{}}
\bibinfo{title}{Switching the magnetic configuration of a spin valve by current-induced domain wall motion}
\bibinfo{journal}{\textit{J. Appl. Phys.}} \textbf{\bibinfo{volume}{92}},
\bibinfo{pages}{4825} (\bibinfo{year}{2002}).  

\bibitem[{\citenamefont{Klaui et~al.}(2003)\citenamefont{Klaui, Vaz, Bland,
Wernsdorfer, Faini, Cambril and Heyderman}}]{Klaui:APL:2003}
\bibinfo{author}{\bibnamefont{Klaui}, \bibfnamefont{M.}}
\bibinfo{author}{\bibnamefont{\textit{el al.}},~\bibfnamefont{}}
\bibinfo{title}{Domain wall motion induced by spin polarized currents in ferromagnetic ring structures}
\bibinfo{journal}{\textit{Appl. Phys. Lett.}} \textbf{\bibinfo{volume}{83}},
\bibinfo{pages}{105-107} (\bibinfo{year}{2003}).    

\bibitem[{\citenamefont{Grollier et~al.}(2003)\citenamefont{Grollier, Boulenc, Cros,
Hamzic, Vaures, Fert and Faini}}]{Grollier:APL:2003}
\bibinfo{author}{\bibnamefont{Grollier}, \bibfnamefont{J.}}
\bibinfo{author}{\bibnamefont{\textit{el al.}},~\bibfnamefont{}}
\bibinfo{title}{Switching a spin valve back and forth by current-induced domain wall motion}
\bibinfo{journal}{\textit{Appl. Phys. Lett.}} \textbf{\bibinfo{volume}{83}},
\bibinfo{pages}{509} (\bibinfo{year}{2003}).  

\bibitem[{\citenamefont{Baibich et~al.}(1988)\citenamefont{Baibich, Broto, Fert, Nguyen Van Dau, Petroff, Etienne, Creuzet, Friederich, Chazelas}}]{Baibich:PRL:1988}
\bibinfo{author}{\bibnamefont{Baibich}, \bibfnamefont{M.N.}}
\bibinfo{author}{\bibnamefont{\textit{el al.}},~\bibfnamefont{}}
\bibinfo{title}{Giant Magnetoresistance of (001)Fe/(001)Cr Magnetic Superlattices}
\bibinfo{journal}{\textit{Phys. Rev. Lett.}} \textbf{\bibinfo{volume}{61}},
\bibinfo{pages}{2472} (\bibinfo{year}{1988}). 

\bibitem[{\citenamefont{Thiaville et~al.}(2004)\citenamefont{Thiaville, Nakatani, Miltat,
Vernier}}]{Thiaville:JAP:2004}
\bibinfo{author}{\bibnamefont{Thiaville}, \bibfnamefont{A.}}
\bibinfo{author}{\bibnamefont{Nakatani},~\bibfnamefont{Y.}}
\bibinfo{author}{\bibnamefont{Miltat},~\bibfnamefont{J.}},
\bibnamefont{\&} 
\bibinfo{author}{\bibnamefont{Vernier},~\bibfnamefont{N.}},
\bibinfo{title}{Domain wall motion by spin-polarized current: a micromagnetic study}
\bibinfo{journal}{\textit{J. Appl. Phys.}} \textbf{\bibinfo{volume}{95}},
\bibinfo{pages}{7049} (\bibinfo{year}{2004}). 

\bibitem[{\citenamefont{Zhang et~al.}(2004)\citenamefont{Zhang, Li}}]{Zhang:PRL:2004}
\bibinfo{author}{\bibnamefont{Zhang}, \bibfnamefont{S.}}
\bibnamefont{\&} 
\bibinfo{author}{\bibnamefont{Li},~\bibfnamefont{L.}},
\bibinfo{title}{Roles of Nonequilibrium Conduction Electrons on the Magnetization Dynamics of Ferromagnets}
\bibinfo{journal}{\textit{Phys. Rev. Lett.}} \textbf{\bibinfo{volume}{93}},
\bibinfo{pages}{127204} (\bibinfo{year}{2004}). 

\bibitem[{\citenamefont{Thiaville et~al.}(2005)\citenamefont{Thiaville, Nakatani, Miltat,
Vernier}}]{Thiaville:EP:2005}
\bibinfo{author}{\bibnamefont{Thiaville}, \bibfnamefont{A.}}
\bibinfo{author}{\bibnamefont{Nakatani},~\bibfnamefont{Y.}}
\bibinfo{author}{\bibnamefont{Miltat},~\bibfnamefont{J.}},
\bibnamefont{\&} 
\bibinfo{author}{\bibnamefont{Suzuki},~\bibfnamefont{Y.}},
\bibinfo{title}{Micromagnetic understanding of current-driven domain wall motion in patterned nanowires}
\bibinfo{journal}{\textit{Europhys. Lett.}} \textbf{\bibinfo{volume}{69}},
\bibinfo{pages}{990} (\bibinfo{year}{2005}). 

\bibitem[{\citenamefont{Tatara et~al.}(2004)\citenamefont{Tatara, Kohno}}]{Tatara:PRL:2004}
\bibinfo{author}{\bibnamefont{Tatara}, \bibfnamefont{G.}}
\bibnamefont{\&} 
\bibinfo{author}{\bibnamefont{Kohno},~\bibfnamefont{H.}},
\bibinfo{title}{Theory of Current-Driven Domain Wall Motion: Spin Transfer versus Momentum Transfer}
\bibinfo{journal}{\textit{Phys. Rev. Lett.}} \textbf{\bibinfo{volume}{92}},
\bibinfo{pages}{086601} (\bibinfo{year}{2004}). 

\bibitem[{\citenamefont{Xiao et~al.}(2006)\citenamefont{Xiao, Zangwill, Stiles}}]{Xiao:PRB:2006}
\bibinfo{author}{\bibnamefont{Xiao}, \bibfnamefont{J.}}
\bibinfo{author}{\bibnamefont{Zangwill}, \bibfnamefont{A.}}
\bibnamefont{\&} 
\bibinfo{author}{\bibnamefont{Stiles},~\bibfnamefont{M.D.}},
\bibinfo{title}{Spin-transfer torque for continuously variable magnetization}
\bibinfo{journal}{\textit{Phys. Rev. B}} \textbf{\bibinfo{volume}{73}},
\bibinfo{pages}{054428} (\bibinfo{year}{2006}). 

\bibitem[{\citenamefont{Kohno et~al.}(2006)\citenamefont{Kohno, Tatara, Shibata}}]{Kohno:JPSJ:2006}
\bibinfo{author}{\bibnamefont{Kohno}, \bibfnamefont{H.}}
\bibinfo{author}{\bibnamefont{Tatara}, \bibfnamefont{G.}}
\bibnamefont{\&} 
\bibinfo{author}{\bibnamefont{Shibata},~\bibfnamefont{J.}},
\bibinfo{title}{Microscopic Calculation of Spin Torques in Disordered Ferromagnets}
\bibinfo{journal}{\textit{J. Phys. Soc. Jpn}} \textbf{\bibinfo{volume}{75}},
\bibinfo{pages}{113706} (\bibinfo{year}{2006}).

\bibitem[{\citenamefont{Tserkovnyak et~al.}(2006)\citenamefont{Tserkovnyak, Skadsem, Brataas, Bauer}}]{Tserkovnyak:PRB:2006}
\bibinfo{author}{\bibnamefont{Tserkovnyak}, \bibfnamefont{Y.}}
\bibinfo{author}{\bibnamefont{Skadsem}, \bibfnamefont{H.J.}}
\bibinfo{author}{\bibnamefont{Braatas}, \bibfnamefont{A.}}
\bibnamefont{\&} 
\bibinfo{author}{\bibnamefont{Bauer},~\bibfnamefont{G.E.W.}},
\bibinfo{title}{Current-induced magnetization dynamics in disordered itinerant ferromagnets}
\bibinfo{journal}{\textit{Phys. Rev. B}} \textbf{\bibinfo{volume}{74}},
\bibinfo{pages}{144405} (\bibinfo{year}{2006}).

\bibitem[{\citenamefont{Garate et~al.}(2009)\citenamefont{Garate, Gilmore, Stiles, MacDonald}}]{Garate:PRB:2009}
\bibinfo{author}{\bibnamefont{Garate}, \bibfnamefont{I.}}
\bibinfo{author}{\bibnamefont{Gilmore}, \bibfnamefont{K.}}
\bibinfo{author}{\bibnamefont{Stiles}, \bibfnamefont{M.D.}}
\bibnamefont{\&} 
\bibinfo{author}{\bibnamefont{MacDonald},~\bibfnamefont{A.H.}},
\bibinfo{title}{Nonadiabatic spin-transfer torque in real materials}
\bibinfo{journal}{\textit{Phys. Rev. B}} \textbf{\bibinfo{volume}{79}},
\bibinfo{pages}{104416} (\bibinfo{year}{2009}). 

\bibitem[{\citenamefont{Meier et~al.}(2007)\citenamefont{Meier, Bolte, Eiselt, Kruger, Kim, Fischer}}]{Meier:PRL:2007}
\bibinfo{author}{\bibnamefont{Meier}, \bibfnamefont{G.}}
\bibinfo{author}{\bibnamefont{\textit{el al.}},~\bibfnamefont{}}
\bibinfo{title}{Direct Imaging of Stochastic Domain-Wall Motion Driven by Nanosecond Current Pulses}
\bibinfo{journal}{\textit{Phys. Rev. Lett.}} \textbf{\bibinfo{volume}{98}},
\bibinfo{pages}{187202} (\bibinfo{year}{2007}). 

\bibitem[{\citenamefont{Thomas et~al.}(2006)\citenamefont{Thomas, Hayashi, Jiang, Moriya, Rettner, Parkin}}]{Thomas:Nature:2006}
\bibinfo{author}{\bibnamefont{Thomas}, \bibfnamefont{L.}}
\bibinfo{author}{\bibnamefont{\textit{el al.}},~\bibfnamefont{}}
\bibinfo{title}{Oscillatory dependence of current-driven magnetic domain wall motion on current pulse length}
\bibinfo{journal}{\textit{Nature}} \textbf{\bibinfo{volume}{447}},
\bibinfo{pages}{197} (\bibinfo{year}{2006}). 

\bibitem[{\citenamefont{Hayashi et~al.}(2008)\citenamefont{Hayashi, Thomas, Rettner, Moriya, Parkin}}]{Hayashi:APL:2008}
\bibinfo{author}{\bibnamefont{Hayashi}, \bibfnamefont{M.}}
\bibinfo{author}{\bibnamefont{\textit{el al.}},~\bibfnamefont{}}
\bibinfo{title}{Dynamics of domain wall depinning driven by a combination of direct and pulsed currents}
\bibinfo{journal}{\textit{Appl. Phys. Lett.}} \textbf{\bibinfo{volume}{92}},
\bibinfo{pages}{162503} (\bibinfo{year}{2008}). 

\bibitem[{\citenamefont{Lepadatu et~al.}(2009)\citenamefont{Lepadatu, Vanhaverbeke, Atkinson, Allenspach, Marrows}}]{Lepadatu:PRL:2009}
\bibinfo{author}{\bibnamefont{Lepadatu}, \bibfnamefont{S.}}
\bibinfo{author}{\bibnamefont{\textit{el al.}},~\bibfnamefont{}}
\bibinfo{title}{Dependence of Domain-Wall Depinning Threshold Current on Pinning Profile}
\bibinfo{journal}{\textit{Phys. Rev. Lett.}} \textbf{\bibinfo{volume}{102}},
\bibinfo{pages}{127203} (\bibinfo{year}{2009}). 

\bibitem[{\citenamefont{Boulle et~al.}(2008)\citenamefont{Boulle, Kimling, Warnicke, Klaui, Rudiger, Malinowski,  Swagten, Koopmans, Ulysse, Faini}}]{Boulle:PRL:2008}
\bibinfo{author}{\bibnamefont{Boulle}, \bibfnamefont{O.}}
\bibinfo{author}{\bibnamefont{\textit{el al.}},~\bibfnamefont{}}
\bibinfo{title}{Nonadiabatic Spin Transfer Torque in High Anisotropy Magnetic Nanowires with Narrow Domain Walls}
\bibinfo{journal}{\textit{Phys. Rev. Lett.}} \textbf{\bibinfo{volume}{101}},
\bibinfo{pages}{216601} (\bibinfo{year}{2008}). 

\bibitem[{\citenamefont{Miron et~al.}(2009)\citenamefont{ Miron,  Zermatten,  Gaudin,  Auffret,  Rodmacq,  Schuhl}}]{Miron:PRL:2009}
\bibinfo{author}{\bibnamefont{Miron}, \bibfnamefont{I. M.}}
\bibinfo{author}{\bibnamefont{\textit{el al.}},~\bibfnamefont{}}
\bibinfo{title}{Domain Wall Spin Torquemeter}
\bibinfo{journal}{\textit{Phys. Rev. Lett.}} \textbf{\bibinfo{volume}{102}},
\bibinfo{pages}{137202} (\bibinfo{year}{2009}). 

\bibitem[{\citenamefont{Miron et~al.}(2010)\citenamefont{Miron, Gaudin, Auffret, Rodmacq, Schuhl, Pizzini, Vogel,  Gambardella}}]{Miron:Nature:2010}
\bibinfo{author}{\bibnamefont{Miron}, \bibfnamefont{I. M.}}
\bibinfo{author}{\bibnamefont{\textit{el al.}},~\bibfnamefont{}}
\bibinfo{title}{Current-driven spin torque induced by the Rashba effect in a ferromagnetic metal layer}
\bibinfo{journal}{\textit{Nature Mat.}} \textbf{\bibinfo{volume}{9}},
\bibinfo{pages}{230} (\bibinfo{year}{2010}). 

\bibitem[{\citenamefont{Chauleau et~al.}(2010)\citenamefont{Chauleau, Weil, Thiaville, Miltat}}]{Chauleau:PRB:2010}
\bibinfo{author}{\bibnamefont{Chauleau}, \bibfnamefont{J.-Y.}}
\bibinfo{author}{\bibnamefont{Weil}, \bibfnamefont{R.}}
\bibinfo{author}{\bibnamefont{Thiaville}, \bibfnamefont{A.}}
\bibnamefont{\&} 
\bibinfo{author}{\bibnamefont{Miltat},~\bibfnamefont{J.}},
\bibinfo{title}{Magnetic domain walls displacement : automotion vs. spin-transfer torque}
\bibinfo{journal}{\textit{Phys. Rev. B}} \textbf{\bibinfo{volume}{82}},
\bibinfo{pages}{214414} (\bibinfo{year}{2010}). 

\bibitem[{\citenamefont{Burrowes et~al.}(2009)\citenamefont{Burrowes, Mihai, Ravelosona, Kim, Chappert, Vila, Marty, Samson, Garcia-Sanchez, Buda-Prejbeanu, Tudosa, Fullerton, Attane}}]{Burrowes:Nature:2009}
\bibinfo{author}{\bibnamefont{Burrowes}, \bibfnamefont{C.}}
\bibinfo{author}{\bibnamefont{\textit{el al.}},~\bibfnamefont{}}
\bibinfo{title}{Non-adiabatic spin-torques in narrow magnetic domain walls }
\bibinfo{journal}{\textit{Nature Phys.}} \textbf{\bibinfo{volume}{6}},
\bibinfo{pages}{19} (\bibinfo{year}{2009}). 

\bibitem[{\citenamefont{Eltschka et~al.}(2010)\citenamefont{Eltschka, Wotzel, Rhensius, Krzyk, Nowak, Klaui, Kasama, Dunin-Borkowski Heyderman, van Driel, Duine}}]{Eltschka:PRL:2010}
\bibinfo{author}{\bibnamefont{Eltschka}, \bibfnamefont{M.}}
\bibinfo{author}{\bibnamefont{\textit{el al.}},~\bibfnamefont{}}
\bibinfo{title}{Non-adiabatic spin torque investigated using thermally activated magnetic domain wall dynamics}
\bibinfo{journal}{\textit{Phys. Rev. Lett.}} \textbf{\bibinfo{volume}{105}},
\bibinfo{pages}{056601}  (\bibinfo{year}{2010}). 

\bibitem[{\citenamefont{Parkin et~al.}(2008)\citenamefont{Parkin, Hayashi, Thomas}}]{Parkin:Science:2008}
\bibinfo{author}{\bibnamefont{Parkin}, \bibfnamefont{S.S.P.}},
\bibinfo{author}{\bibnamefont{Hayashi}, \bibfnamefont{M.}},
\bibnamefont{\&}
\bibinfo{author}{\bibnamefont{Thomas}, \bibfnamefont{L.}},
\bibinfo{title}{Magnetic Domain-Wall Racetrack Memory} \bibinfo{journal}{\textit{Science}} \textbf{\bibinfo{volume}{320}},
\bibinfo{pages}{190-194} (\bibinfo{year}{2008}).

\bibitem[{\citenamefont{Yuasa et~al.}(2005)\citenamefont{Yuasa, Nagahama, Fukushima, Suzuki and Ando}}]{Yuasa:NatMat:2004}
\bibinfo{author}{\bibnamefont{Yuasa}, \bibfnamefont{S.}},
\bibinfo{author}{\bibnamefont{Nagahama}, \bibfnamefont{T.}},
\bibinfo{author}{\bibnamefont{Fukushima}, \bibfnamefont{A.}},
\bibinfo{author}{\bibnamefont{Suzuki}, \bibfnamefont{Y.}}
\bibnamefont{\&}
\bibinfo{author}{\bibnamefont{Ando}, \bibfnamefont{K.}},
\bibinfo{title}{Giant room-temperature magnetoresistance in single-crystal Fe/MgO/Fe magnetic tunnel junctions}
\bibinfo{journal}{\textit{Nature Mater.}}
\textbf{\bibinfo{volume}{3}}, \bibinfo{pages}{868-871} (\bibinfo{year}{2004}).

\bibitem[{\citenamefont{Parkin et~al.}(2005)\citenamefont{Parkin, Kaiser, Panchula, Rice, Hughes, Samant and Yang}}]{Parkin:NatMat:2004}
\bibinfo{author}{\bibnamefont{Parkin}, \bibfnamefont{S.S.P.}},
\bibinfo{author}{\bibnamefont{\textit{el al.}},~\bibfnamefont{}}
\bibinfo{title}{Giant tunnelling magnetoresistance at room temperature with MgO (100) tunnel barriers}
\bibinfo{journal}{\textit{Nature Mater.}}
\textbf{\bibinfo{volume}{3}}, \bibinfo{pages}{862-867} (\bibinfo{year}{2004}).

\bibitem[{\citenamefont{Cros}(2004)}]{Cros:Patent:2004}
\bibinfo{author}{\bibnamefont{Cros}, \bibfnamefont{V.}},
\bibinfo{author}{\bibnamefont{Grollier}, \bibfnamefont{J.}},
\bibinfo{author}{\bibnamefont{Munoz Sanchez}, \bibfnamefont{M.}},
\bibinfo{author}{\bibnamefont{Fert}, \bibfnamefont{A.}},
\bibnamefont{\&} 
\bibinfo{author}{\bibnamefont{Nguyen van Dau}, \bibfnamefont{F.}},
\bibinfo{title}{Spintronic device with control by domain wall displacement induced by a current of spin polarized carriers}  
\bibinfo{journal}{\textit{Patent}},
\bibinfo{pages}{FR 2004/04-13338, WO 2006/06 4022 A1, US2009/0273421 A1} (\bibinfo{year}{2004}).

\bibitem[{\citenamefont{NEC}(2009)}]{NEC}
\bibinfo{author}{\bibnamefont{Fukami}, \bibfnamefont{S.}},
\bibinfo{author}{\bibnamefont{\textit{el al.}},~\bibfnamefont{}}
\bibinfo{title}{Low-current perpendicular domain wall motion cell for scalable high-speed MRAM}  
\bibinfo{journal}{\textit{VLSI Technology, 2009 Symposium on, Honolulu, HI}},
\bibinfo{pages}{230 - 231} (\bibinfo{year}{2009}).

\bibitem[{\citenamefont{Strukov et~al.}(2008)\citenamefont{Strukov, Snider, Stewart, Williams}}]{Strukov:Nature:2008}
\bibinfo{author}{\bibnamefont{Strukov}, \bibfnamefont{D.B.}}
\bibinfo{author}{\bibnamefont{Snider},~\bibfnamefont{G.S.}}
\bibinfo{author}{\bibnamefont{Stewart},~\bibfnamefont{D.R.}},
\bibnamefont{\&} 
\bibinfo{author}{\bibnamefont{Williams},~\bibfnamefont{R.S.}},  
\bibinfo{title}{The missing memristor found}
\bibinfo{journal}{\textit{Nature}} \textbf{\bibinfo{volume}{453}},
\bibinfo{pages}{80} (\bibinfo{year}{2008}). 

\bibitem[{\citenamefont{Chua}(1971)\citenamefont{Chua}}]{Chua:IEEE:1971}
\bibinfo{author}{\bibnamefont{Chua},~\bibfnamefont{L.O.}},  
\bibinfo{title}{Memristor - The missing circuit element}
\bibinfo{journal}{\textit{IEEE Trans. Circuit Theory}} \textbf{\bibinfo{volume}{18}},
\bibinfo{pages}{507} (\bibinfo{year}{1971}). 

\bibitem[{\citenamefont{Yang et~al.}(2008)\citenamefont{Yang, Pickett, Li, Ohlberg, Stewart, Williams}}]{Yang:Nature:2008}
\bibinfo{author}{\bibnamefont{Yang}, \bibfnamefont{J.J.}}
\bibinfo{author}{\bibnamefont{\textit{el al.}},~\bibfnamefont{}}
\bibinfo{title}{Memristive switching mechanism for metal/oxide/metal nanodevices}
\bibinfo{journal}{\textit{Nature Nano.}} \textbf{\bibinfo{volume}{3}},
\bibinfo{pages}{429} (\bibinfo{year}{2008}). 

\bibitem[{\citenamefont{Yamaguchi et~al.}(2004)\citenamefont{Yamaguchi, Ono, Nasu, Miyake, Mibu,  Shinjo}}]{Yamaguchi:PRL:2004}
\bibinfo{author}{\bibnamefont{Yamaguchi}, \bibfnamefont{A.}}
\bibinfo{author}{\bibnamefont{\textit{el al.}},~\bibfnamefont{}}  
\bibinfo{title}{Real-Space Observation of Current-Driven Domain Wall Motion in Submicron Magnetic Wires}
\bibinfo{journal}{\textit{Phys. Rev. Lett.}} \textbf{\bibinfo{volume}{92}},
\bibinfo{pages}{077205} (\bibinfo{year}{2004}). 

\bibitem[{\citenamefont{Yamanouchi et~al.}(2006)\citenamefont{Yamanouchi, Chiba, Matsukura, Dietl, Ohno}}]{Yamanouchi:PRL:2006}
\bibinfo{author}{\bibnamefont{Yamanouchi}, \bibfnamefont{M.}}
\bibinfo{author}{\bibnamefont{Chiba},~\bibfnamefont{D.}}
\bibinfo{author}{\bibnamefont{Matsukura},~\bibfnamefont{F.}}
\bibinfo{author}{\bibnamefont{Dietl},~\bibfnamefont{T.}},
\bibnamefont{\&} 
\bibinfo{author}{\bibnamefont{Ohno},~\bibfnamefont{H.}},  
\bibinfo{title}{Velocity of Domain-Wall Motion Induced by Electrical Current in the Ferromagnetic Semiconductor (Ga,Mn)As}
\bibinfo{journal}{\textit{Phys. Rev. Lett.}} \textbf{\bibinfo{volume}{96}},
\bibinfo{pages}{096601} (\bibinfo{year}{2006}). 

\bibitem[{\citenamefont{Wang et~al.}(2009)\citenamefont{Wang, Chen, Xi, Li, Dimitrov}}]{Wang:IEEE:2009}
\bibinfo{author}{\bibnamefont{Wang}, \bibfnamefont{X.}}
\bibinfo{author}{\bibnamefont{Chen},~\bibfnamefont{Y.}}
\bibinfo{author}{\bibnamefont{Xi},~\bibfnamefont{H.}},
\bibinfo{author}{\bibnamefont{Li},~\bibfnamefont{H.}},
\bibnamefont{\&} 
\bibinfo{author}{\bibnamefont{Dimitrov},~\bibfnamefont{D.}},  
\bibinfo{title}{Spintronic Memristor Through Spin-Torque-Induced Magnetization Motion}
\bibinfo{journal}{\textit{IEEE Electron Device Letters}} \textbf{\bibinfo{volume}{30}},
\bibinfo{pages}{294} (\bibinfo{year}{2009}). 

\bibitem[{\citenamefont{Grollier et~al.}(2010)\citenamefont{Grollier, Cros, Nguyen Van dau}}]{Grollier:patent:2010}
\bibinfo{author}{\bibnamefont{Grollier}, \bibfnamefont{J.}}
\bibinfo{author}{\bibnamefont{Cros},~\bibfnamefont{V.}}
\bibnamefont{\&} 
\bibinfo{author}{\bibnamefont{Nguyen Van dau},~\bibfnamefont{F.}},  
\bibinfo{title}{Memristor device with resistance adjustable by moving a magnetic wall by spin transfer and use of said memristor in a neural network}
\bibinfo{journal}{\textit{Patent}} \textbf{\bibinfo{volume}{WO/2010/125181}},
(\bibinfo{year}{2010}). 

\bibitem[{\citenamefont{Slonczewski}(2005)\citenamefont{Slonczewski}}]{Slonczewski:PRB:2005}
\bibinfo{author}{\bibnamefont{Slonczewski},~\bibfnamefont{J.C.}},  
\bibinfo{title}{Currents, torques, and polarization factors in magnetic tunnel junctions}
\bibinfo{journal}{\textit{Phys. Rev. B}} \textbf{\bibinfo{volume}{71}},
\bibinfo{pages}{024411} (\bibinfo{year}{2005}).  

\bibitem[{\citenamefont{Theodonis et~al.}(2006)\citenamefont{Theodonis, Kioussis, Kalitsov, Chshiev, Butler}}]{Theodonis:PRL:2006}
\bibinfo{author}{\bibnamefont{Theodonis}, \bibfnamefont{I.}}
\bibinfo{author}{\bibnamefont{Kioussis},~\bibfnamefont{N.}}
\bibinfo{author}{\bibnamefont{Kalitsov},~\bibfnamefont{A.}},
\bibinfo{author}{\bibnamefont{Chshiev},~\bibfnamefont{M.}},
\bibnamefont{\&} 
\bibinfo{author}{\bibnamefont{Butler},~\bibfnamefont{W.H.}},  
\bibinfo{title}{Anomalous Bias Dependence of Spin Torque in Magnetic Tunnel Junctions}
\bibinfo{journal}{\textit{Phys. Rev. Lett.}} \textbf{\bibinfo{volume}{97}},
\bibinfo{pages}{237205} (\bibinfo{year}{2006}). 

\bibitem[{\citenamefont{Ravelosona et~al.}(2006)\citenamefont{Ravelosona, Mangin, Lemaho, Katine, Terris, Fullerton}}]{Ravelosona:PRL:2006}
\bibinfo{author}{\bibnamefont{Ravelosona}, \bibfnamefont{D.}},
\bibinfo{author}{\bibnamefont{\textit{el al.}},~\bibfnamefont{}}
\bibinfo{title}{Domain Wall Creation in Nanostructures Driven by a Spin-Polarized Current} \bibinfo{journal}{\textit{Phys. Rev. Lett.}} \textbf{\bibinfo{volume}{96}},
\bibinfo{pages}{186604} (\bibinfo{year}{2006}).
\bibitem[{\citenamefont{Rebei et~al.}(2006)\citenamefont{Rebei, Mryasov }}]{Rebei:PRB:2006}
\bibinfo{author}{\bibnamefont{Rebei}, \bibfnamefont{A.}},
\bibnamefont{\&}
\bibinfo{author}{\bibnamefont{Mryasov}, \bibfnamefont{O.}},
\bibinfo{title}{Dynamics of a trapped domain wall in a spin-valve nanostructure with current perpendicular to the plane}  
\bibinfo{journal}{\textit{Phys. Rev. B}} \textbf{\bibinfo{volume}{74}},
\bibinfo{pages}{014412} (\bibinfo{year}{2006}).
\bibitem[{\citenamefont{Lou et~al.}(2008)\citenamefont{Lou, Gao, Dimitrov, Tang}}]{Lou:APL:2008}
\bibinfo{author}{\bibnamefont{Lou}, \bibfnamefont{X.}},
\bibinfo{author}{\bibnamefont{Gao}, \bibfnamefont{Z.}},
\bibinfo{author}{\bibnamefont{Dimitrov}, \bibfnamefont{D.V.}},
\bibnamefont{\&}
\bibinfo{author}{\bibnamefont{Tang}, \bibfnamefont{M.X.}},
\bibinfo{title}{Demonstration of multilevel cell spin transfer switching in MgO magnetic tunnel junctions} \bibinfo{journal}{\textit{Appl. Phys. Lett.}} \textbf{\bibinfo{volume}{93}},
\bibinfo{pages}{242502} (\bibinfo{year}{2008}).
\bibitem[{\citenamefont{Boone et~al.}(2010)\citenamefont{Boone, Katine, Carey, Childress, Cheng, Krivorotov}}]{Boone:PRL:2010}
\bibinfo{author}{\bibnamefont{Boone}, \bibfnamefont{C.T.}},
\bibinfo{author}{\bibnamefont{\textit{el al.}},~\bibfnamefont{}}
\bibinfo{title}{Rapid Domain Wall Motion in Permalloy Nanowires Excited by a Spin-Polarized Current Applied Perpendicular to the Nanowire} \bibinfo{journal}{\textit{Phys. Rev. Lett.}} \textbf{\bibinfo{volume}{104}},
\bibinfo{pages}{097203} (\bibinfo{year}{2010}).

\bibitem[{\citenamefont{Khvalkovskiy et~al.}(2009)\citenamefont{Khvalkovskiy, Zvezdin, Gorbunov, Cros, Grollier, Fert, Zvezdin }}]{Khvalkovskiy:PRL:2009}
\bibinfo{author}{\bibnamefont{Khvalkovskiy}, \bibfnamefont{K.V.}},
\bibinfo{author}{\bibnamefont{\textit{el al.}},~\bibfnamefont{}}
\bibinfo{title}{High domain wall velocities due to spin currents perpendicular to the plane}  
\bibinfo{journal}{\textit{Phys. Rev. Lett.}} \textbf{\bibinfo{volume}{102}},
\bibinfo{pages}{067206} (\bibinfo{year}{2009}).

\bibitem[{\citenamefont{Stiles et~al.}(2002)\citenamefont{Stiles, Zangwill}}]{Stiles:PRB:2002}
\bibinfo{author}{\bibnamefont{Stiles}, \bibfnamefont{M.D.}},
\bibnamefont{\&}
\bibinfo{author}{\bibnamefont{Zangwill}, \bibfnamefont{A.}},
\bibinfo{title}{Anatomy of spin-transfer torque}  
\bibinfo{journal}{\textit{Phys. Rev. B}} \textbf{\bibinfo{volume}{66}},
\bibinfo{pages}{014407} (\bibinfo{year}{2002}).

\bibitem[{\citenamefont{Xia et~al.}(2002)\citenamefont{Xia, Kelly, Bauer, Brataas, Turek}}]{Xia:PRB:2002}
\bibinfo{author}{\bibnamefont{Xia}, \bibfnamefont{K.}},
\bibinfo{author}{\bibnamefont{Kelly}, \bibfnamefont{P.J.}},
\bibinfo{author}{\bibnamefont{Bauer}, \bibfnamefont{G.E.W.}},
\bibinfo{author}{\bibnamefont{Brataas}, \bibfnamefont{A.}},
\bibnamefont{\&}
\bibinfo{author}{\bibnamefont{Turek}, \bibfnamefont{I.}},
\bibinfo{title}{Spin torques in ferromagnetic/normal-metal structures}  
\bibinfo{journal}{\textit{Phys. Rev. B}} \textbf{\bibinfo{volume}{65}},
\bibinfo{pages}{220401(R)} (\bibinfo{year}{2002}).

\bibitem[{\citenamefont{Sankey et~al.}(2007)\citenamefont{Sankey, Cui, Sun, Slonczewski, Buhrman, Ralph }}]{Sankey:Nature:2007}
\bibinfo{author}{\bibnamefont{Sankey}, \bibfnamefont{J.C.}},
\bibinfo{author}{\bibnamefont{\textit{el al.}},~\bibfnamefont{}}
\bibinfo{title}{Measurement of the spin-transfer-torque vector in magnetic tunnel junctions}  
\bibinfo{journal}{\textit{Nature Physics}} \textbf{\bibinfo{volume}{4}},
\bibinfo{pages}{67-71} (\bibinfo{year}{2007}).

\bibitem[{\citenamefont{Kubota et~al.}(2007)\citenamefont{Kubota, Fukushima, Yakushiji, Nagahama, Yuasa, Ando, Maehara, Nagamine, Tsunekawa, Djayaprawira, Watanabe, Suzuki }}]{Kubota:Nature:2007}
\bibinfo{author}{\bibnamefont{Kubota}, \bibfnamefont{H.}},
\bibinfo{author}{\bibnamefont{\textit{el al.}},~\bibfnamefont{}}
\bibinfo{title}{Quantitative measurement of voltage dependence of spin-transfer torque in MgO-based magnetic tunnel junctions}   \bibinfo{journal}{\textit{Nature Physics}} \textbf{\bibinfo{volume}{4}},
\bibinfo{pages}{37-41} (\bibinfo{year}{2007}).

\bibitem[{\citenamefont{Oh et~al.}(2009)\citenamefont{Oh, Park, Manchon, Chshiev, Han, Lee, Lee, Nam, Jo, Kong, Dieny, Lee }}]{Oh:Nature:2009}
\bibinfo{author}{\bibnamefont{Oh}, \bibfnamefont{S.-C.}},
\bibinfo{author}{\bibnamefont{\textit{el al.}},~\bibfnamefont{}}
\bibinfo{title}{Bias-voltage dependence of perpendicular spin-transfer torque in asymmetric MgO-based magnetic tunnel junctions}   \bibinfo{journal}{\textit{Nature Physics}} \textbf{\bibinfo{volume}{5}},
\bibinfo{pages}{898-902} (\bibinfo{year}{2009}).

\bibitem[{\citenamefont{Tang et~al.}(2010)\citenamefont{Tang, Kioussis, Kalitsov, Butler, Car}}]{Tang:PRB:2010}
\bibinfo{author}{\bibnamefont{Tang}, \bibfnamefont{Y.-H.}},
\bibinfo{author}{\bibnamefont{Kioussis}, \bibfnamefont{N.}},
\bibinfo{author}{\bibnamefont{Kalitsov}, \bibfnamefont{A.}},
\bibinfo{author}{\bibnamefont{Butler}, \bibfnamefont{W.H.}},
\bibnamefont{\&}
\bibinfo{author}{\bibnamefont{Car}, \bibfnamefont{R.}},
\bibinfo{title}{Influence of asymmetry on bias behavior of spin torque}   \bibinfo{journal}{\textit{Phys. Rev. B}} \textbf{\bibinfo{volume}{81}},
\bibinfo{pages}{054437} (\bibinfo{year}{2010}).

\end{thebibliography}
\end{document}